\documentclass[sigconf]{acmart}
\usepackage[ruled,vlined]{algorithm2e}
\usepackage{multirow}
\usepackage{subcaption}
\newtheorem{theorem}{Theorem}

\AtBeginDocument{%
  \providecommand\BibTeX{{%
    \normalfont B\kern-0.5em{\scshape i\kern-0.25em b}\kern-0.8em\TeX}}}

\setcopyright{acmcopyright}
\copyrightyear{2020}
\acmYear{2020}
\acmDOI{10.1145/1122445.1122456}

\acmConference[San Diego '2020]{San Diego '2020: ACM SIGKDD Conference on Knowledge Discovery and Data Mining}{Aug 21--26, 2020}{San Diego, CA}
\acmBooktitle{San Diego '2020: ACM SIGKDD Conference on Knowledge Discovery and Data Mining,
  Aug 21--26, 2020, San Diego, CA}
\acmPrice{15.00}
\acmISBN{978-1-4503-XXXX-X/18/06}

\begin{document}

\title{Cascade-BGNN: Toward Efficient Self-supervised Representation Learning on Large-scale Bipartite Graphs}
\author{Chaoyang He}
\authornote{Both authors contributed equally to this research.}
\email{chaoyang.he@usc.edu}
\orcid{1234-5678-9012}
\author{Tian Xie}
\authornotemark[1]
\email{xiet@usc.edu}
\affiliation{%
  \institution{University of Southern California}
  \streetaddress{}
  \city{Los Angeles}
  \state{California}
  \country{USA}
  \postcode{}
}

\author{Yu Rong}
\affiliation{%
  \institution{Tencent AI Lab}
  \streetaddress{}
  \city{Shenzhen}
  \country{China}}
\email{yu.rong@hotmail.com}

\author{Wenbing Huang}
\affiliation{%
  \institution{Tencent AI Lab}
  \streetaddress{}
  \city{Shenzhen}
  \country{China}}
\email{hwenbing@126.com}


\author{Junzhou Huang}
\affiliation{%
  \institution{Tencent AI Lab}
  \streetaddress{}
  \city{Shenzhen}
  \country{China}}
\email{jzhuang@uta.edu}

\author{Xiang Ren}
\affiliation{%
  \institution{University of Southern California}
  \streetaddress{}
  \city{Los Angeles}
  \state{California}
  \country{USA}
}
\email{xiangren@usc.edu}

\author{Cyrus Shahabi}
\affiliation{%
  \institution{University of Southern California}
  \streetaddress{}
  \city{Los Angeles}
  \state{California}
  \country{USA}
}
\email{shahabi@usc.edu}

\renewcommand{\shortauthors}{Trovato and Tobin, et al.}

\begin{abstract}
Bipartite graphs have been used to represent data relationships in many data-mining applications such as in E-commerce recommendation systems. Since learning in graph space is more complicated than in Euclidian space, recent studies have extensively utilized neural nets to effectively and efficiently embed a graph's nodes into a multidimensional space. However, this embedding method has not yet been applied to large-scale bipartite graphs. Existing techniques either cannot be scaled to large-scale bipartite graphs that have limited labels or cannot exploit the unique structure of bipartite graphs, which have distinct node features in two domains. Thus, we propose Cascade Bipartite Graph Neural Networks, Cascade-BGNN, a novel node representation learning for bipartite graphs that is domain-consistent, self-supervised, and efficient. To efficiently aggregate information both across and within the two partitions of a bipartite graph, BGNN utilizes a customized Inter-domain Message Passing (IDMP) and Intra-domain Alignment (IDA), which is our adaptation of adversarial learning, for message aggregation across and within partitions, respectively. BGNN is trained in a self-supervised manner.
Moreover, we formulate a multi-layer BGNN in a cascaded training manner to enable multi-hop relationship modeling while improving training efficiency. Extensive experiments on several datasets of varying scales verify the effectiveness and efficiency of BGNN over baselines. Our design is further affirmed through theoretical analysis for domain alignment. The scalability of BGNN is additionally verified through its demonstrated rapid training speed and low memory cost over a large-scale real-world bipartite graph \footnote{Our code is open-sourced at \url{https://github.com/chaoyanghe/bipartite-graph-learning}}.
\end{abstract}




\keywords{graph neural networks, self-supervised learning, graph embedding, representation learning}


\maketitle
\section{Introduction}
\label{sec:introduction}
\begin{figure}[t]
    \centering
    \includegraphics[width=0.90\linewidth]{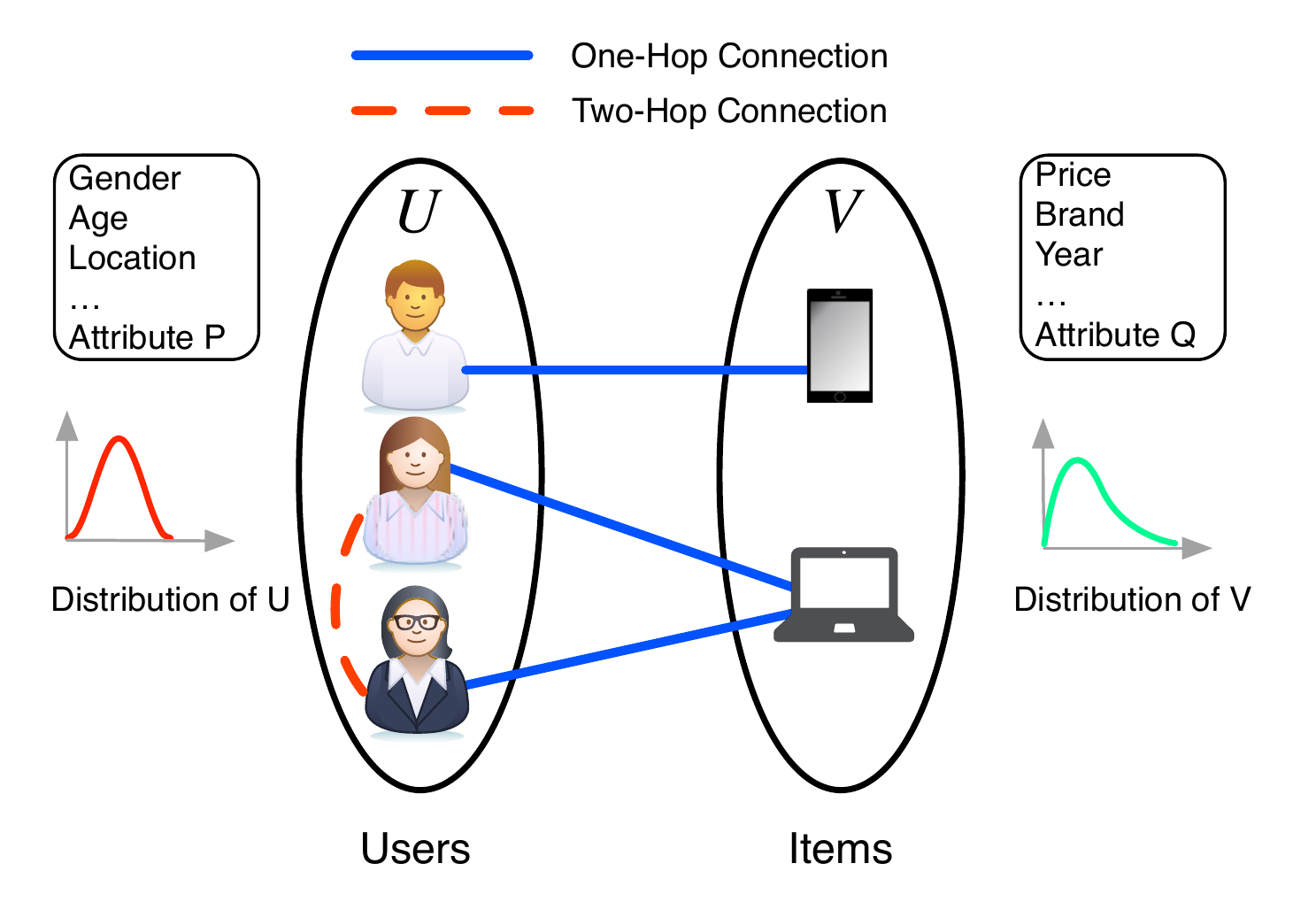}
    \vskip -0.1 in
    \caption{A Bipartite Graph in E-commerce system.}
    \vskip 0.1 in
    \label{fig: bipartite_graphs}
\end{figure}

Graphs have been used to capture and represent complex structural relationships among data items in various domains, including drug discovery \cite{you_graph_2018,jin_junction_2018}, social networks analysis \cite{wang_structural_2016,qiu_deepinf:_2018}, and visual understanding \cite{ferrari_graph_2018,wan_representation_2018}. Amongst their varying forms, bipartite graphs are prevalent in data mining applications. A bipartite graph (Fig. \ref{fig: bipartite_graphs}) is a graph whose vertices are divided into two independent partitions such that every edge connects nodes from one partition to the other. For example, in an e-commerce recommendation system, the two distinct partitions are represented by users and products, and an edge from a member from one partition to a member of the other represents the user purchasing the product \cite{linden_amazon.com_2003}. 
The ability to utilize information from the graphical structure, such as node features in the two distinct partitions and topology information, plays an important role in the accuracy and effectiveness of services and tasks, such as classification, prediction, and recommendation. 
\begin{figure*}[ht]
  \includegraphics[width=\textwidth]{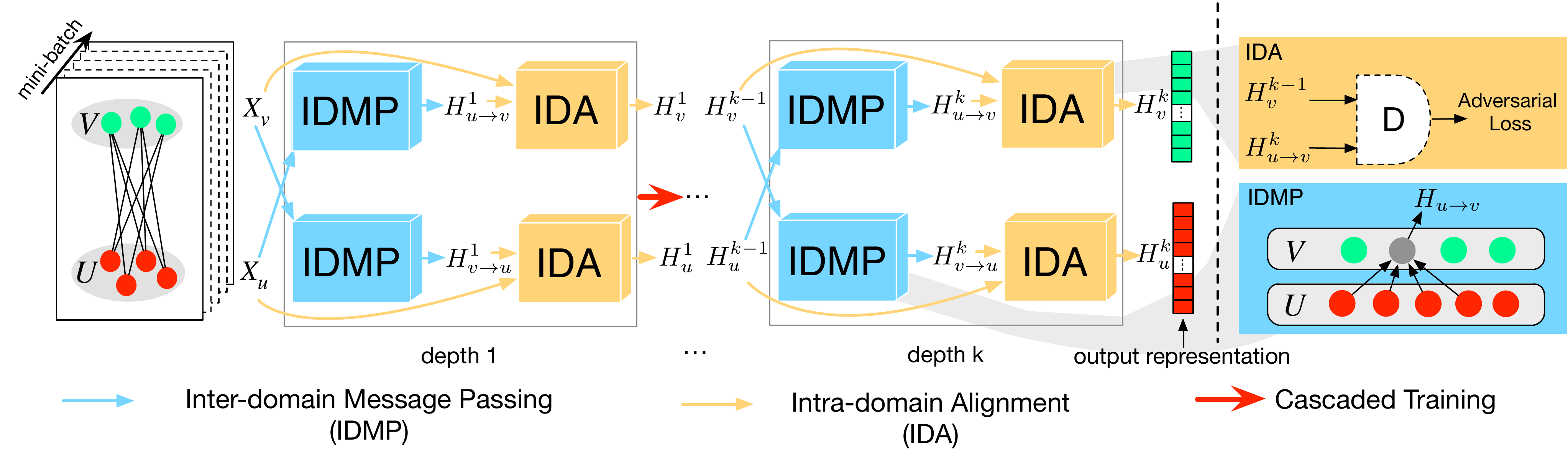}
  \vskip -0.1 in
  \caption{Illustration of Cascade-BGNN. Given the inputs of two domains $X_u$ and $X_v$, we obtain their self-supervised representation as $H_u$ and $H_v$ via inter-domain message passing and intra-domain distribution alignment. To enable multi-hop neighbor information aggregation, we stack multiple layers to formulate a deep BGNN whose layers are trained in a cascading manner.}
  \vskip -0.2 in
  \label{fig:BGNN}
\end{figure*}

Graph representation learning, also called graph embedding, is a machine learning paradigm that aims to solve the above problem by mapping structural information into a low-dimensional vector space which can then be used to improve the performance of downstream tasks \cite{wu2019comprehensive}. Early classical works include random walk-based methods (DeepWalk \cite{perozzi_deepwalk:_2014}, Node2Vec\cite{grover_node2vec:_2016}), where only graph topology and node relations are embedded as vectors. 
In light of the rapid advancements of deep learning, Graph Neural Networks (GNNs) \cite{scarselli_graph_2009} have exhibited tremendous
progress in representation learning for generic graphs (GraphSAGE~\cite{hamilton_inductive_2017}, AS-GCN \cite{huang_adaptive_2018}).
In general, GNNs recursively update each node's feature by aggregating its neighbors through message passing, by which not only the patterns of graph topology but also node features are captured. However, GNNs cannot embed the rich information contained in distinct node features from two domains and topology information into a single node presentation, especially for an extremely large-scale bipartite graph without sufficient node-wise labels. For example, in the e-commerce system (Fig. 1), to precisely represent a user, hidden product attributes should also be extracted and incorporated into the final embedding vector because it can express a user's taste and interest, which is useful in accurately classifying users. 

This paper studies the above problem of node representation learning on large-scale bipartite graphs in a self-supervised manner. There are two main challenges. The first challenge with bipartite graphs is that features of nodes in \textit{each partition of a bipartite graph may follow different distributions in distinct feature domains} (e.g., in Fig. \ref{fig: bipartite_graphs}, users and products have different attributes). Therefore, representing bipartite graphs as generic graphs to leverage neighbor message passing, as is typically done with GNNs, fails to exploit the extra knowledge from two distinct node domains. One can deal with two different partitions for each domain by converting two-hop neighbors to one-hop homogeneous connections within the same domain (note that each node and its two-hop neighbors are in the same partition). This approach is clearly unable to exploit the feature correlations across the two partitions. Alternatively, one can treat bipartite graphs as heterogeneous networks and use random walk-based methods such as \textit{Metapath2Vec} \cite{dong_metapath2vec:_2017}. 
However, \textit{Metapath2Vec} does not integrate node features into the embedding process. Additionally, a meta-path in a bipartite graph can be equated to a naive unbiased random walk in homogeneous graphs. 

The second challenge is that \textit{the limited label issue} and \textit{computational inefficiency} become non-trivial when scaling up extremely large bipartite graphs. Limited labels prohibit supervised learning. For example, an e-commerce bipartite graph contains billions of users, and labeling every user requires tremendous effort, and is thus impractical. Normally, only an extremely small fraction of the billions of users is manually labeled, which is insufficient for supervised learning. Some unsupervised learning methods have been proposed to address this problem \cite{hu2019pre,liu2019n}. However, they either do not apply to bipartite graphs or are inefficient in scalability.
In terms of computational inefficiency, scaling up is difficult. Although sampling methods, such as GraphSAGE\cite{hamilton_inductive_2017}, AS-GCN\cite{huang_adaptive_2018} and FastGCN\cite{chen_fastgcn:_2018}), have been proposed to deal with the scalability issue (uncontrollable neighborhood expansion across layers), when applied to bipartite, they are only able to propagate through multiple layers in the same feature domain by connecting indirect two-hop nodes, which significantly increases the edge number. Thus, their training speed and memory costs are still unsatisfied.

To address the limitations of existing methods, we propose cascade-BGNN (Bipartite Graph Neural Networks), a highly efficient graph neural network framework that we have developed and deployed in production. Cascade-BGNN outperforms current competitive baselines in terms of both effectiveness and efficiency in a large-scale bipartite graph where nodes and edges are on the order of millions or billions \footnote{Because of privacy and regulatory restrictions, in this paper, we only release a million-level dataset for the experimental demonstration.} - a graph that is 1000 times larger than typical applications of GNNs. These advantages are due to the three key designs of our Cascade-BGNN.

\textbf{Domain-consistent.} To represent the node features within different domains into a single representation, BGNN consists of two central operations: \emph{inter-domain message passing} (IDMP) and \emph{intra-domain alignment} (IDA). As illustrated in Fig. ~\ref{fig:BGNN}, two node domains are represented as $U$ and $V$, respectively. In each layer (depth) of BGNN, we formulate two simultaneous directions of information flow, one from $U$ to $V$ and the other from $V$ to $U$, each of which is equipped with different weight filters. By aggregating information from another domain through the connected edges, IDMP can attain an inter-domain representation for each node, which is then fused with the raw feature in each node itself. For domain fusion, we propose an intra-domain alignment technique to minimize the divergence between raw features and the inter-domain representation by using adversarial models \cite{goodfellow_generative_2014,pan_adversarially_2018}, a tool that has been applied successfully for distribution matching. 

\textbf{Self-Supervised}. Most notably, BGNN is tailored for a limited-label setting. BGNN is a self-supervised learning framework, which is a form of unsupervised learning in which the data itself provides supervision. One may argue that a straightforward method to fuse the two domains is to concatenate the input feature of each node with its corresponding inter-domain representation as an enhanced output. Nevertheless, such a method requires using the final labels as supervised signals for the training process, which is not feasible in the limited label setting. 

\textbf{Efficient in Large-Scale}. Our model and training method co-design elegantly improve the scalability. We design a \textit{cascaded training} method that allows for multi-stage training without supervision. That is, the training of the upper layer (depth $k+1$ in Fig. \ref{fig:BGNN}) begins only after the lower one (depth $k$ in Fig. \ref{fig:BGNN}) has been trained completely. 
Cascaded training is clearly more memory-efficient than conventional end-to-end training method since it does not require restoration of all intermediate activation maps of neural layers. Additionally, in cascaded training, the domain shift in lower layers (\emph{i.e.} the discrepancy between two domain features, which always exists during the early training phase) is not passed to higher ones. In contrast, in end-to-end training, this type of error accumulates as the depth increases. More details are discussed in \textsection~\ref{sec:cascaded}.

We provide theoretical analysis and empirical verification to demonstrate the advantages of Cascade-BGNN. Our theoretical analysis proves that domain alignment is able to make the representation distribution closer. Such distribution approaching causes information from one domain to be incorporated into the other. In our experiments, we contrast the performance of our algorithm with several unsupervised representation learning baselines: Node2Vec~\cite{grover_node2vec:_2016}, VGAE~\cite{kipf_variational_2016}, GraphSAGE~\cite{hamilton_inductive_2017}, and AS-GCN \cite{huang_adaptive_2018}. We use a large-scale bipartite graph dataset from the Tencent Platform and also construct three synthesized datasets based on the citation networks Cora, Citeseer, and PubMed \cite{sen_collective_2008}. For all benchmarks, BGNN outperforms other competitive baselines in terms of both effectiveness and efficiency, with a higher classification accuracy, faster training speed, and lower memory cost.

To our knowledge, this is the largest self-supervised representation learning framework for bipartite graphs. Our source code and large dataset are released for reproducibility and future research.

\section{Problem Statement}
We define \textit{Bipartite Graphs} as follows: Let $G=(U, V, E)$ be a bipartite graph (as illustrated in Fig. 1), where $U$ and $V$ denote the set of the two domains of vertices (nodes). $u_i$ and $v_j$ denote the $i$-th and $j$-th vertex in $U$ and $V$, respectively, where $i=1,2,...,M$ and $j=1,2,...,N$. There are only inter-domain edges, which are defined as $E\subseteq{U\times{V}}$. $e_{ij}$ represents the edge between $u_i$ and $v_j$. The incidence matrix for set $U$ is $B_u\in{\mathbb{R}^{M\times{N}}}$ and $B_v\in{\mathbb{R}^{N\times{M}}}$ for set $V$. $B_{u(i,j)} = 1$ if $e_{ij}\in{E}$, and $B_{u(i,j)}=0 \text { if } e_{i j} \notin E$. 
The features of two sets of nodes can be formulated as $X_u$ and $X_v$, respectively, where $X_u\in\mathbb{R}^{M\times{P}}$ is a feature matrix with $x_{u(i)}\in\mathbb{R}^P$ representing the feature vector of node $u_i$, and $X_v\in\mathbb{R}^{N\times{Q}}$ is similarly defined.

A core assumption is that the number of nodes and edges in bipartite graphs might be extremely large (on the order of millions or billions) and that there are limited node-wise labels.

Our work focuses on designing a self-supervised node representation learning model that can exploit both topology information and distinct node features from two domains to improve the accuracy of downstream tasks (e.g., classification). 

\section{Proposed model: BGNN}
\label{sec:BGNN}
In this section, we introduce Cascade-BGNN, Cascade Bipartite Graph Neural Networks, a self-supervised representation learning framework for large-scale bipartite graphs. We also summarize the overall algorithm and provide a theoretical analysis.

\paragraph{Overall Framework}
In general, \textit{bipartite graph representation learning} aims to learn the embedding representation $H_{u}\in\mathbb{R}^{P^{'}}$ and $H_{v}\in\mathbb{R}^{Q^{'}}$ for nodes in group $U$ and $V$, respectively. Let $f_{emb}$ be a general bipartite graph embedding model with parameters $\theta$. In order to embed distinct node features $X_u$ and $X_v$, the representation of $H_{u}$ and $H_{v}$ is defined as follows:
\begin{eqnarray}
\label{Eq:f_emb}
H_u, H_v &=& f_{emb}(X_u, B_u, X_v, B_v; \theta)
\end{eqnarray}
The entire architecture of $f_{emb}$ is illustrated in Fig. \ref{fig:BGNN}. There are three key designs within. The first is inter-domain message passing (IDMP), which is represented in blue in Fig. \ref{fig:BGNN}. Its goal is for one domain to aggregate information from the other domain through the connected edges. Formally, it can be expressed as:
\begin{eqnarray}
\label{Eq:IDMP-u}
H_{v \rightarrow u} &=& f_{u}(X_v, B_u; \theta_u) \\
\label{Eq:IDMP-v}
H_{u \rightarrow v} &=& f_{v}(X_u, B_v; \theta_v)
\end{eqnarray}
where $f_u$ and $f_v$ are the IDMP function for these two domains respectively, $H_{v \rightarrow u}$ (resp. $H_{u \rightarrow v}$) represents aggregated information flowing from $V$ (resp. $U$) to $U$ (resp. $V$). More details of IDMP are provided in \textsection~\ref{sec:IDMP}.

Once the aggregated features from the opposite domain $H_{v \rightarrow u}$, $H_{u \rightarrow v}$ are attached, we use intra-domain alignment (IDA) to fuse these two distinct features into a single representation. IDA is represented by orange in Fig. \ref{fig:BGNN}. Formally, we express it as:
\begin{eqnarray}
\label{Eq:IDA-u}
Loss_u = L_{adv}(H_{v \rightarrow u}, X_u) \\
\label{Eq:IDA-v}
Loss_v = L_{adv}(H_{u \rightarrow v}, X_v)
\end{eqnarray}
where $L_{adv}$ is specified as an adversarial loss.  After one layer training through minimization of Eq.~\eqref{Eq:IDA-u}-\eqref{Eq:IDA-v} in a self-supervised manner, we obtain the representation of the two domains $H_u^1$ and $H_v^1$. Further explanation for IDA is provided in~\textsection~\ref{subsec:IDMP-Adv}. We also provide a theoretical analysis in~\textsection~\ref{sec:theory} to explain our design choice of the domain alignment.

The embedding of $H_u^1$ (resp. $H_v^1$) merely captures the one-hop topology structure of $B_u$ (resp. $B_v$) as well as feature information from $X_u$ and $X_v$. As presented in previous works~\cite{kipf_semi-supervised_2016,hamilton_inductive_2017}, the one-hop aggregation does not sufficiently characterize diverse graph structures, hence a multi-hop mechanism is required. Instead of leveraging the typical end-to-end training method, this paper develops a cascaded training method to drive multi-hop message passing. Additionally, cascaded training is far more scalable. We will detail the cascaded training in~\textsection~\ref{subsec: cascaded}.

\refstepcounter{secnumdepth}
\subsection{Inter-Domain Message Passing (IDMP)}
\label{sec:IDMP}
Formally, the adjacency matrix of a bipartite graph is 
\begin{equation}
    A = \begin{pmatrix} 0_{u, u} & B_u \\ B_v & 0_{v, v}\end{pmatrix}
\label{label:adj_matrix}
\end{equation}
where $B_u$ and $B_v$ are incidence matrices for two partitions, respectively. For stability, we normalize $B_u$ as $\hat{B}_u=D_u^{-1} B_u$, where $D_u$ is the degree matrix of $B_u$. Similar normalization is done for $B_v$. The IDMP process is defined as 
\begin{equation}
\label{label:IDMP}
\left\{
    \begin{array}{lr}
    H_{v \rightarrow u}^{(k)} = \sigma(\hat{B}_u H_v^{(k)} W_u^{(k)})     &  \\
    H_{u \rightarrow v}^{(k)} = \sigma(\hat{B}_v H_u^{(k)} W_v^{(k)})     & 
    \end{array}
\right.
\end{equation}
where $\sigma$ denotes an activation function, such as ReLU, $H_{v \rightarrow u}^{k}\in{\mathbb{R}^{M\times{Q}^{'}}}$ (resp. $H_{u \rightarrow v}^{(k)}\in{\mathbb{R}^{N\times{P}^{'}}}$) are hidden features of the nodes in set $U$ (resp. $V$) aggregated from the features in $V$ (resp. $U$), and $k$ indicates the depth index (note that when $k=0$, $H^{(0)}_{u} = X_{u}, H^{(0)}_{v} = X_{v}$ are actually input features).

As we can see from Eq.\ref{label:IDMP}, there are two distinctions between IDMP and conventional GCNs \cite{hamilton_inductive_2017}: 1. IDMP only performs aggregation on each node's neighbor nodes without involving the node itself, while conventional GCN methods usually consider the self-loop computation; 2. The propagation is only one-hop-neighbor aware. The first distinction motivates us to further design an intra-domain alignment to take the self-input features into account, while the second one leads to our design in the cascaded training approach which enables multi-hop modeling and supports efficient training.

\subsection{Intra-Domain Alignment (IDA)}
\label{subsec:IDMP-Adv}
\textit{We introduce IDA from the perspective of the domain $U$}. We design two types of alignment losses to align $H_{v \rightarrow u}$ with $H_u$. Our first alignment employs adversarial learning \cite{goodfellow_generative_2014,ganin_domain-adversarial_2015}. A \textit{discriminator} is trained to discriminate between vectors randomly sampled from $H_{v \rightarrow u} $ and $H_u$. Conversely, IDMP (Inter-Domain Message Passing) is trained as a \textit{generator} to prevent the discriminator from predicting accurately. As a result, this
becomes a two-player min-max game, where the discriminator aims to maximize the ability to identify two distinct feature representations, and IDMP aims to prevent the discriminator from doing so by making the encoded representation $H_{v \rightarrow u}$ (source) to approach $H_u$ (target). After training, they will reach a Nash equilibrium, successfully aligned $H_{v \rightarrow u}$ and $H_u$.

\textbf{Discriminator objective:} we define parameters of IDA discriminator as $\theta$ and IDMP generator as $\phi$. We denote $P_{\theta, \phi} (\text{source} = 1|h)$ as the probability that the input feature vector $h$ is from the source domain $H_{v \rightarrow u}$. Conversely, $\text{source} = 0$ signifies that $h$ is from the target domain $H_u$. The discriminator loss function is as follows:
\begin{equation}
\begin{aligned}
L_{D}(\theta|\phi) = & \frac{1}{M}\sum_{i=1}^{M} \log P_{\theta, \phi} (\text{source} = 0|h_{u(i)})\\
-&\frac{1}{N}\sum_{i=1}^{N} \log P_{\theta, \phi} (\text{source} = 1|h_{v\rightarrow{u}(i)})
\end{aligned}
\end{equation}

\textbf{Generator objective:} in the generative setting, IDMP is trained to align the encoded representation $H_{v \rightarrow u}$ (source) to $H_u$ (target) so that the discriminator is unable to distinguish them:
\begin{equation}
\begin{aligned}
L_{G}(\phi|\theta) =& \frac{1}{N}\sum_{i=1}^{N} \log P_{\theta, \phi} (\text{source} = 0|h_{v\rightarrow{u}(i)}) \\
\end{aligned} 
\end{equation}

During training, the discriminator and the generator are trained successively with gradient updates to optimize the two networks, respectively. In experiment \textsection~\ref{exp:ida}, when using this method for domain alignment, we call our model BGNN-Adv. 

Another intuitive approach is to utilize multi-layer perceptron (MLP) as IDA to project node features $H_u$ and IDMP output $H_{v \rightarrow u} $ into the same feature space. This approach is relatively straight-forward, but we can compare it with BGNN-Adv to judge whether or not adversarial learning can align the domain effectively. Formally, we define the loss function for one set $U$ as
\begin{equation}
    {L}_u = ||\text{MLP}(H^{(i)}_{v \rightarrow u})-H^{(i-1)}_u||_F
\end{equation}
which is symmetric for set $V$. The multi-layer perceptron takes the IDMP output as its input and minimizes it with original features in a Frobenius norm. In the experiment \textsection~\ref{exp:ida}, we term our model BGNN-MLP when using MLP as IDA. We show that BGNN-Adv outperforms BGNN-MLP on the classification task. 

\subsection{Cascaded Training: Towards Efficient BGNN}
\begin{figure}[h]
\centering
\begin{subfigure}{0.4\textwidth}
    {{\includegraphics[width=1\textwidth]{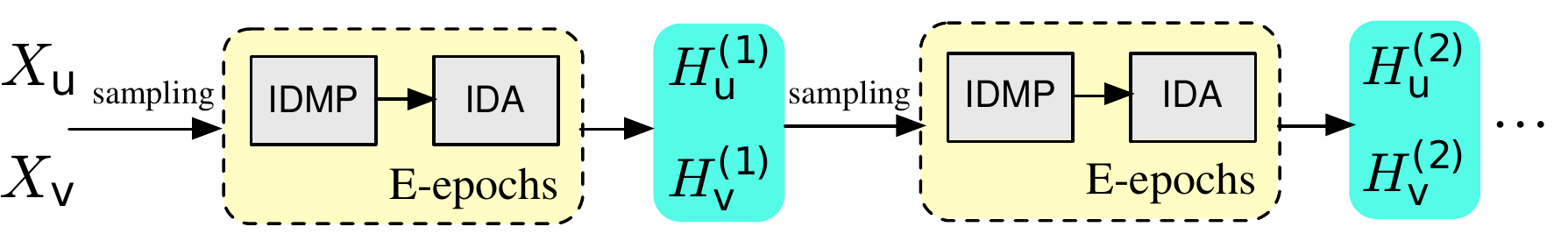}}}
    \caption{Our Cascaded Training}
\end{subfigure}

\begin{subfigure}{0.4\textwidth}
    {{\includegraphics[width=1\textwidth]{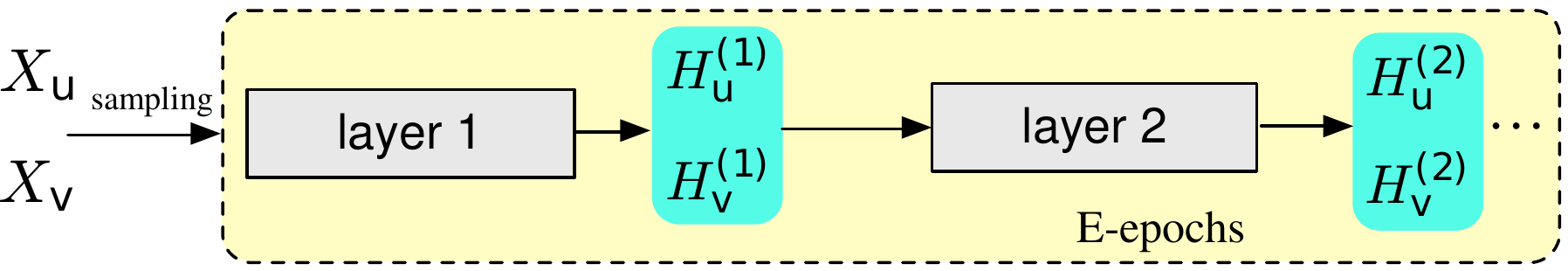}}}
    \caption{Traditional End-to-end Training}
\end{subfigure}
\caption{Comparison of cascaded training with traditional end-to-end training.}
\label{fig:Sampling}
\end{figure}

\label{subsec: cascaded}
We present the cascaded training design for our proposed BGNN. In Fig. \ref{fig:Sampling}, we depict a detailed diagram to illustrate our cascaded training process in comparison with the conventional end-to-end training paradigm. In cascaded training, we regard one depth (layer) training as training on a basic BGNN block (IDMP with IDA). Each depth completes its training with one-hop embedding in $E$ epochs. Then, its trained embedding is used as the input for later training. This process is also illustrated in Fig. \ref{fig:BGNN}. Cascaded training is in contrast to the conventional end-to-end training paradigm (shown in Fig. \ref{fig:Sampling}), which propagates through multiple depths for a fixed number of hops to train the final embedding.

We argue that cascaded training can embed information from multi-hops in bipartite graphs, similarly to how GNN-based methods perform on general graphs. Additionally, cascaded training is more memory-efficient and also requires less training time. System-wise advantages demonstrate our design choice:
 
\textbf{Only one depth training is alive}. This indicates that our method requires a significantly lower memory cost. In Fig. \ref{fig:Sampling}, each depth in a cascaded architecture takes the final embedding from the previous depth as its input. This indicates that we can destroy model instance (release unused memory) in the previous depth and only keep one depth training alive throughout the entire training process. 

\textbf{Avoid uncontrollable neighborhood expansion}. On a large-scale graph, the uncontrollable neighborhood expansion of each node layer by layer leads to a low computational speed and high memory cost. Sampling methods like GraphSAGE \cite{hamilton_inductive_2017} and AS-GCN \cite{huang_adaptive_2018} have been proposed to deal with this issue by reducing the number of neighborhood nodes in each layer. Compared to these methods, since the propagation of our architecture only happens one-hop, the neighborhood expansion issue is avoided, which consequently speeds up the training and reduces memory cost.


\textbf{Robust in hyper-parameter tuning}. Our cascaded architecture is robust in that it can be easily trained with minimal hyper-parameter tuning effort. By using cascaded training, the domain shift (i.e, the discrepancy between two domain features, which always exists during the early training phase) in lower depths will not be passed to higher ones; while in end-to-end training this kind of error accumulates as the depth increases.

\textbf{With out statistical performance sacrifice}. Notably, system-wise optimization does not sacrifice the statistical performance: multi-hop topology information can also be preserved by cascaded training, and it simultaneously reduces the memory cost and training time on large-scale bipartite graphs. 

We verified the advantages of cascaded training in~\textsection~\ref{sec:cascaded} and ~\textsection~\ref{subsec:system}.

\subsection{Algorithm}
\begin{algorithm}
\caption{Cascade-BGNN algorithm}
\SetAlgoLined
\KwIn{Graph $G(U, V, E)$; input features $\left\{X_u, X_v\right\}$}
\KwOut{Node representation $Z_u$ and $Z_v$} 
$H^0_u \gets X_u$; $H^0_v \gets X_v$\\
\For{$k = 1, ...K$}{
 \For{e in epochs}{
 Sampling batches $(h_u^{(k)}, h_v^{(k)})$ from $(H_u^{(k)}, H_v^{(k)})$ \\
 \For{$h^{(k)}_{u}$, $h^{(k)}_{v}$ as batches of input}{
  $h^{(k+1)}_{u} \gets \text{IDA}^{(k)}(h^{(k)}_{u}, \underbrace{\text{IDMP}^{(k)}(h_v^{(k)})}_{h_{v \rightarrow u}^{(k)}})$\;
  $h^{(k+1)}_{v} \gets \text{IDA}^{(k)}(h^{(k)}_{v}, \underbrace{\text{IDMP}^{(k)}(h_u^{(k)})}_{h_{u \rightarrow v}^{(k)}})$\;
  }
 }
 Save($H_u^{(k+1)}$, $H_v^{(k+1)}$) for $(k+1)$th depth training \\
 Release($\text{IDMP}^{(k)}$, $\text{IDA}^{(k)}$, $H_u^{k}$, $H_v^{k}$)
}
$Z_u \gets H_u^K$; $Z_v \gets H_v^K$
\label{alg:BGNN-algorithm}
\end{algorithm}
We summarize our overall implementation framework for Cascade-BGNN model in Algorithm \ref{alg:BGNN-algorithm}, which is consistent with Fig. \ref{fig:BGNN} and Fig.  \ref{fig:Sampling}. The processes for set $U$ and $V$ are symmetric. Each step in the outmost loop proceeds as follows, where $k$ represents the current layer and $H^{(k)}_u, H^{(k)}_v$ are hidden representations in layer $k$. For every epoch, sampling is conducted on these hidden representations to get a mini-batch as input. After several epochs of training, the embedding representation of depth $k$ can be learned and saved for $k+1$ layer training. The $k$th layer model instance and unused memory are released. The final representation can be extracted in the last layer $K$, which can then be used for downstream tasks. The time complexity per epoch for Cascade-BGNN is fixed at $O(|E|)$ ($|E|$ denotes the number of edges), where there is no neighborhood expansion along with layer (depth) in traditional end-to-end training.

\subsection{Theoretical Analysis}
\label{sec:theory}
In this section, we show that the better the alignment of embedding space in IDA, the better the downstream model \textit{M} performs on the node representation. We show this by theoretically analyzing the relationship between alignment effects and the difference between the loss function value on model \textit{M} on these two domains representations when handling a node classification task. 

We denote the nodes' label by $y$ and its representation vector by $h$. There are two domains, $H_u$ and $H_{v\rightarrow u}$ from output of IDMP. We assume that $M$ outputs the conditional distribution of a node's label $y$ based on its representation vector $h$ and model parameter $\theta$, denoted as $\hat{P}(y|v;\theta)$. This gives the probability that a node has a label given the embedding vector of the node. We can write the loss function of the model $M$ on one domain $H_u$ as $L_{M, u}$:
\begin{equation}
    L_{M, u} = \mathbb{E}(D(\hat{P}_u(y|h;\theta), {P}(y|h)))
\end{equation}
where ${P}(y|h)$ is the ground truth, and $D(P_1, P_2)$ measures the distance of two distributions. $\hat{P}_u(y|h;\theta)$ means that this prediction is based on training data from the $U$ domain. We can rewrite the expectation as:
\begin{equation}
    L_{M, u} = \sum_{h}p_u(h) \cdot D(\hat{P}_u(y|h;\theta), {P}(y|h))
\end{equation}
where $p_u(h)$ is the distribution in the embedding space. Similarly, the performance of the same model based on another domain $H_{v\rightarrow u}$ training data can be measured by the loss function:
\begin{equation}
\begin{aligned}
    L_{M, v\rightarrow u} & = \mathbb{E}(D(\hat{P}_{v\rightarrow u}(y|h;\theta), P(y|h)))\\
    & = \sum_{h}p_{v\rightarrow u}(h) \cdot D(\hat{P}_{v\rightarrow u}(y|h;\theta), {P}(y|h))
\end{aligned}
\end{equation}

We introduce a theorem:
\begin{theorem}
\label{theorem_1}
If following inequalities are satisfied:
\begin{equation}
    D(\hat{P}_{u}(y|h;\theta), \hat{P}_{v\rightarrow u}(y|h;\theta)) < d, \forall h \in H
\label{eq:14}
\end{equation}
\begin{equation}
    \frac{|p_u(h) - p_{v\rightarrow u}(h)|}{p_{v\rightarrow u}(h)} < \epsilon, \forall h \in H
\label{eq:15}
\end{equation}
Then we will have following inequality:
\begin{equation}
    L_{M, u} \leq (1 + \epsilon) L_{M, v\rightarrow u} + d
\label{eq:16}
\end{equation}
\end{theorem}

\textit{The Proof for Theorem 1 can be found in the appendix.}

This theorem shows that by optimizing the representation on the domain $H_{v\rightarrow u}$, we are also improving the representation on the domain $H_u$, which is in the form of achieving a lower loss bounded by $L_{M, v\rightarrow u}$ on the downstream classification model $M$ in equation Eq. \ref{eq:16}. The closer loss of these two domains produces a similar embedding space, which captures information from both domains. Thus, in order to achieve this, we need to make $\epsilon$ and $d$ in Eq. \ref{eq:14} and \ref{eq:15} smaller. In other words, we need to force distribution $p_u(h)$ and $p_{v\rightarrow u}(h)$ to keep close in the same embedding space. Furthermore, if two domains are close in the embedding space, then models trained on these two domains should output similar classification distributions, which guarantees to have a smaller probability distance in Eq. \ref{eq:14}. We achieve the above goals through adversarial learning in BGNN. Consequently, this distribution alignment by IDA is guaranteed to capture information from both domains and result in better representation. 

\section{Experiments}
\label{sec:Experiments}
We design our experiments with the goals of (i) providing a rigorous comparison of the graph representation performance between our BGNN model and state-of-art baselines,  (ii) verifying domain alignment and cascaded training, (iii) evaluating the BGNN efficiency of space and time complexity on a large-scale dataset.
\subsection{Dataset}
In this section, we introduce datasets that we used for experiments and the methods that we prepossessed these datasets.
\begin{table}[h]
    \centering
    \caption{Dataset statistics}
    \begin{tabular}{c|c|cccc}
        \toprule
         \multicolumn{2}{c|}{Dataset} & Tencent & Cora & Citeseer & PubMed \\
         \midrule\midrule
         \multicolumn{2}{c|}{\#Edges}  & 991,734 & 1,802 & 1,000 & 18,782 \\
         \cline{1-6}
         \multirow{2}{*}{\#Nodes} & U & 619,030 & 734 & 613 & 13,424 \\
         \cline{2-6}
         & V & 90,044 & 877 & 510 & 3,435 \\
         \cline{1-6}
         \multirow{2}{*}{\#Features} & U & 8 & 1,433 & 3,703 & 400 \\
         \cline{2-6}
         & V & 16 & 1,000 & 3,000 & 500 \\
         \cline{1-6}
         \multirow{2}{*}{\#Classes} & U & 2 & 7 & 6 & 3 \\
         \cline{2-6}
         & V & N/A & 6 & 6 & 3 \\
        \bottomrule
    \end{tabular}
    \label{table:statistic}
\end{table}

\textbf{Tencent - Social Networks.} This is a large-scale real-world social network represented by a bipartite graph. Nodes in set $U$ are social network users, and nodes in set $V$ are social communities (e.g., a subset of social network users who share the same interests in electrical products may join the same shopping community). Both users and communities are described by dense off-the-shelf feature vectors.  The edge connection between two sets indicates that the user belongs to the community. Note that this dataset provides classification labels for research purposes. In real-world applications, labeling every node is impractical.

\textbf{Cora, Citeseer, PubMed - Citation Networks.} These are \textit{synthetic} bipartite graph datasets that are generated from citation networks (single graph) where documents and citation links between them are treated as nodes and undirected edges, respectively. 

\textbf{Data Preprocessing.} The process that we synthesize the bipartite graphs from citation networks is as follows: We process the Cora, CiteSeer, and PubMed datasets similarly and treat the original graph as an undirected graph. First, we divide the paper documents of each class into two equal-sized subsets. Then, we combine the first half of all classes into the $U$ group and the second half into the $V$ group. We remove some of the features of papers in the $V$ group to introduce heterogeneity between $U$ and $V$.  Lastly, we only keep edges that connected a paper in $U$ group and a paper in $V$ group and remove all other edges to make the graph bipartite. All isolated nodes are removed. Note that the parameters used for preprocessing do not affect the fairness of the comparison: we have verified that the relative ranking of the performance comparison on different baselines will not change with differing split proportions and dimensional heterogeneity.

As the Tencent dataset is already a bipartite graph, there is no need to change the graphical structure. To simplify the data loading process, we maintain the same format as the citation network datasets. The statistics of our datasets are summarized in Table \ref{table:statistic}. 

\subsection{Experimental Settings}
\begin{table*}[ht]
    \centering
    \caption{Prediction results for the four datasets ($F_1$ scores). Results for BGNN unsupervised nodes embedding are shown. (OOM means out of memory)}
    \vspace{-2ex}
    \begin{tabular}{cccccccc}
        \toprule
        & \multicolumn{1}{c}{Tencent} & \multicolumn{2}{c}{Cora} & \multicolumn{2}{c}{Citeseer} & \multicolumn{2}{c}{PubMed}  \\
        \cmidrule(r){2-2}\cmidrule(r){3-4}\cmidrule(r){5-6}\cmidrule(r){7-8} Methods & $F_1$ & Micro $F_1$ & Macro $F_1$ & Micro $F_1$ & Macro $F_1$ & Micro $F_1$ & Macro $F_1$ \\
        \midrule
        Raw features & 0.497 & 0.789 & 0.758 & 0.707 & 0.621 & 0.838 & 0.843 \\ 
        Node2Vec & 0.577 & 0.810 & 0.780 & 0.724 & 0.627  & 0.834 & 0.839  \\ 
        VGAE  & OOM & 0.782 & 0.754 & 0.732  & 0.645 & 0.823 &  0.828 \\ 
        GraphSAGE-GCN  & 0.529  & 0.782  & 0.763 & 0.715  & 0.627 & 0.838 &  0.843 \\
        GraphSAGE-MEAN & 0.580  & 0.823 & 0.801 & 0.748 & 0.665 & 0.838 &  0.843 \\ 
        BGNN-Adv  & \textbf{0.622$\pm$0.017}  & \textbf{0.859$\pm$0.005}  & \textbf{0.831$\pm$0.007} & \textbf{0.768$\pm$0.004}  & \textbf{0.698$\pm$0.005}  & \textbf{0.857$\pm$0.005} &  \textbf{0.860$\pm$0.005} \\ 
        \midrule
        \% gain over feat.  &  25\% & 9\% & 9\% & 9\% & 12\%  & 2\% & 2\% \\
        \bottomrule
    \end{tabular}
    \label{table:performance}
\end{table*}

\textbf{Baselines for comparison.} We mainly compare our BGNN algorithm against four unsupervised node embedding baselines: 
\begin{itemize}
    \item \textbf{Raw features}: This indicates a naive solution in which only raw features are used as input for the classification model, without using any graph structure information incorporated.
    \item \textbf{Node2Vec} \cite{grover_node2vec:_2016}: This approach is an extension of Word2Vec \cite{mikolov_distributed_2013} on graph, which learns a feature representation by simulating biased random walks on the graph. We run Node2Vec on the bipartite graph and then concatenate the node embeddings with their own features.
    \item \textbf{VGAE} \cite{kipf_variational_2016}: This method is based on a variational auto-encoder, where GCN is used as an encoder and a simple inner product as a decoder to embed the nodes into a low-dimensional feature space.
    \item \textbf{GraphSAGE-MEAN}, \textbf{GraphSAGE-GCN}
    \cite{hamilton_inductive_2017,kipf_semi-supervised_2016}: We implement two types of aggregator functions: GCN and MEAN aggregator. Node-wise sampling is used to address the scalability issue. 
    \item \textbf{AS-GCN} \cite{huang_adaptive_2018}: This method uses adaptive sampling between each layer to deal with node explosion in large-scale graphs. Since this method is originally designed for supervised learning, we only compare its scalability in \textsection~\ref{subsec:system}.
\end{itemize}
Note that we do not compare our method with other baselines, e.g. Methpath2vec++, because these models are not tailored to bipartite graphs: they cannot embed distinct features in two domains into a single representation; meta-path in bipartite graphs is the same as unbiased random walk in one of two subgraphs. Therefore, node2vec is adequate to represent them as baselines. 

We cannot directly apply GCN to bipartite graphs due to the inconsistency of the nodes’ feature dimensions in the two bipartite partitions. To make the comparison available, we reconstruct the bipartite graph into two subgraphs, where each only contains nodes from one partition with their two-hops connection through the opposite partition, as shown in Fig. 1. Through this conversion, GCN based methods can be implemented on two partitions, each with the same feature dimensions respectively, but still containing the original connectivity information in bipartite graphs. 

For each baseline model, we follow the open-source implementation from the authors' original paper (in the appendix, we introduced the source code we used). In order to provide a fair comparison, we also tune the hyper-parameters for every baseline and report the best results among them. 
For the adversarial learning in IDA, we use a hyperbolic function as the non-linear activation function in the graph convolution networks. The dropout and L2 regularization are applied to each layer to prevent overfitting. During training, we use mini-batch to reduce the memory and computational cost for the large-scale dataset. We found that the optimal batch size for all four data set is near 500, and it only requires around 3 epochs on each data set to quickly converge to the best result. \textit{More details about parameter settings for each dataset are shown in the appendix}.

Experiments are conducted on a GPU server with 8 Tesla V100 cards. We describe more details about our tailored system for our BGNN in the Appendix.

\subsection{Evaluation Results}
\label{subsec:results}

We evaluated our BGNN results on a classification downstream task with $F_1$ score \cite{hamilton_inductive_2017} which is a popular metric for classification. For binary classification on the Tencent dataset, we report $F_1$ scores. For other multi-classification tasks, we use both micro- and macro-averaged $F_1$ scores.

\textbf{Performance comparison.} From Table \ref{table:performance} we can see that BGNN-Adv achieves the best performance on bipartite graph representation learning. BGNN surpasses other methods on both large and small data sets, suggesting its effectiveness in capturing both inter-domain and intra-domain information. Particularly on the Tencent large-scale bipartite graph, BGNN-Adv achieves a $29\%$ gain over the raw feature baseline. In the PubMed dataset, due to the balanced degree distribution, even incorporating the graph structure information into the model can only marginally improve the accuracy. However, BGNN still achieves the best results among other baselines, which verifies that it also has a better capacity to embed more information on different long-tail datasets and balanced datasets. The overall evaluation proves that Cascade-BGNN, using distinct features and topology information, is effective for graph representation learning.


\subsection{Effect of Distribution Alignment}
\label{exp:ida}
In this section, we conduct ablation studies for the IDA component. For domain fusion, one straightforward method is to concatenate the input feature of each node with its corresponding inter-domain representation as an enhanced feature vector. Nevertheless, such a simple concatenation requires supervising signals, which are not feasible for large-scale graphs that only have very limited node labels. Therefore, to demonstrate the effectiveness of IDA, our ablation studies still focus on unsupervised learning.
\begin{table}[h]
    \centering
    \caption{Comparsion of result without domain alignment with BGNN-Adv (Micro $F_1$ score)}
    \begin{tabular}{ccccc}
        \toprule
        & {Tencent} & {Cora} & {Citeseer} & {PubMed}  \\
        \midrule
        Raw features & 0.497 & 0.789 & 0.707 & 0.838  \\ 
        Feature aggregation & 0.453 & 0.625 & 0.431 & 0.569  \\ 
        BGNN-Adv & \textbf{0.622} & \textbf{0.859} & \textbf{0.768} & \textbf{0.857}  \\ 
        \bottomrule
    \end{tabular}
    \label{table:ablation_study}
\end{table}

\textbf{Comparison with the feature concatenation method.} To show the importance of distribution alignment, we compare BGNN with raw features (Raw features in Table 3) input and feature aggregated from another domain (Feature aggregation in Table 3) without any adversarial training. The result is shown in Table 3. BGNN-Adv significantly outperforms these two naive baselines that have no distribution alignment. Both baselines only contain part of the information from the entire graph, which provides a limited representation. This also corresponds with our theory that merging information from two domains in a bipartite graph will lead to improvement in the final representation.

\begin{table}[h]
    \centering
    \caption{Comparsion of BGNN-MLP with BGNN-Adv (Micro $F_1$ score)}.
    \begin{tabular}{ccccc}
        \toprule
        & {Tencent} & {Cora} & {Citeseer} & {PubMed}  \\
        \midrule
        BGNN-MLP & 0.582 & 0.784 & 0.756 & 0.846  \\ 
        BGNN-Adv & 0.622 & 0.859 & 0.768 & 0.857  \\ 
        \bottomrule
    \end{tabular}
    \label{table:ablation_study}
\end{table}
\textbf{BGNN-Adv v.s. BGNN-MLP.} We replace the adversarial IDA into MLP for self supervision, which we introduced in ~\textsection~\ref{subsec:IDMP-Adv}. Compared to BGNN-Adv, BGNN-MLP shows lower prediction results. The reason is that MLP only performs linear mapping in high dimensions between two domains distribution, which limits its ability in nonlinear type distribution alignment typically seen in high dimensions.

\subsection{Effect of Cascaded Training}
\label{sec:cascaded}
In this section, we verified our cascaded training design by comparing it with end-to-end training and evaluating the results on different amounts of cascaded layers.

\textbf{Self-supervised training loss.} We demonstrate the training loss of cascaded architecture. During each layer of training, we wait until the loss functions converge before continuing to the next layer step. To reduce the memory cost, rather than training two domains simultaneously in each layer, we alternatively select one domain to train the model. 
\begin{figure}[t]
  \centering
  \includegraphics[width=1.0\linewidth]{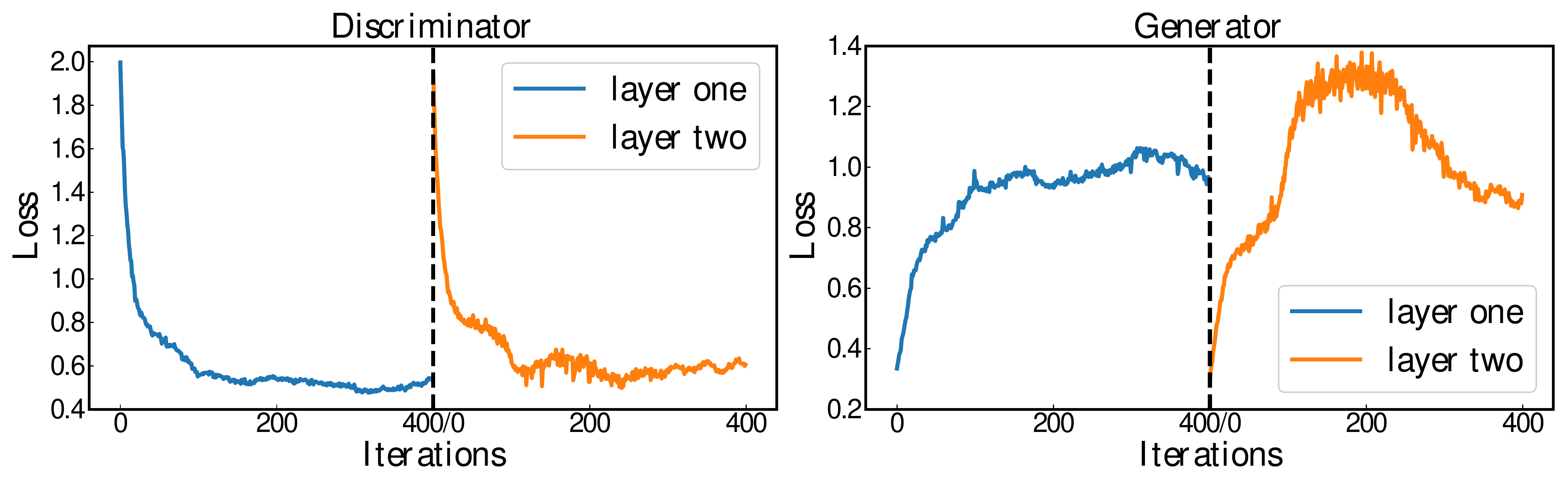}
\caption{Adversarial training loss of cascaded architecture. The $x$ axis denotes iteration numbers in each layer.}
\label{fig:training_loss}
\end{figure}
The adversarial training loss function of BGNN is shown in Fig. \ref{fig:training_loss} where we only plot training losses of the first and second layers due to limited space. The generator and discriminator losses clearly demonstrate that the IDA performs an intra-domain alignment in an adversarial way. 
\begin{figure}[t]
  \centering
  \includegraphics[width=1.0\linewidth]{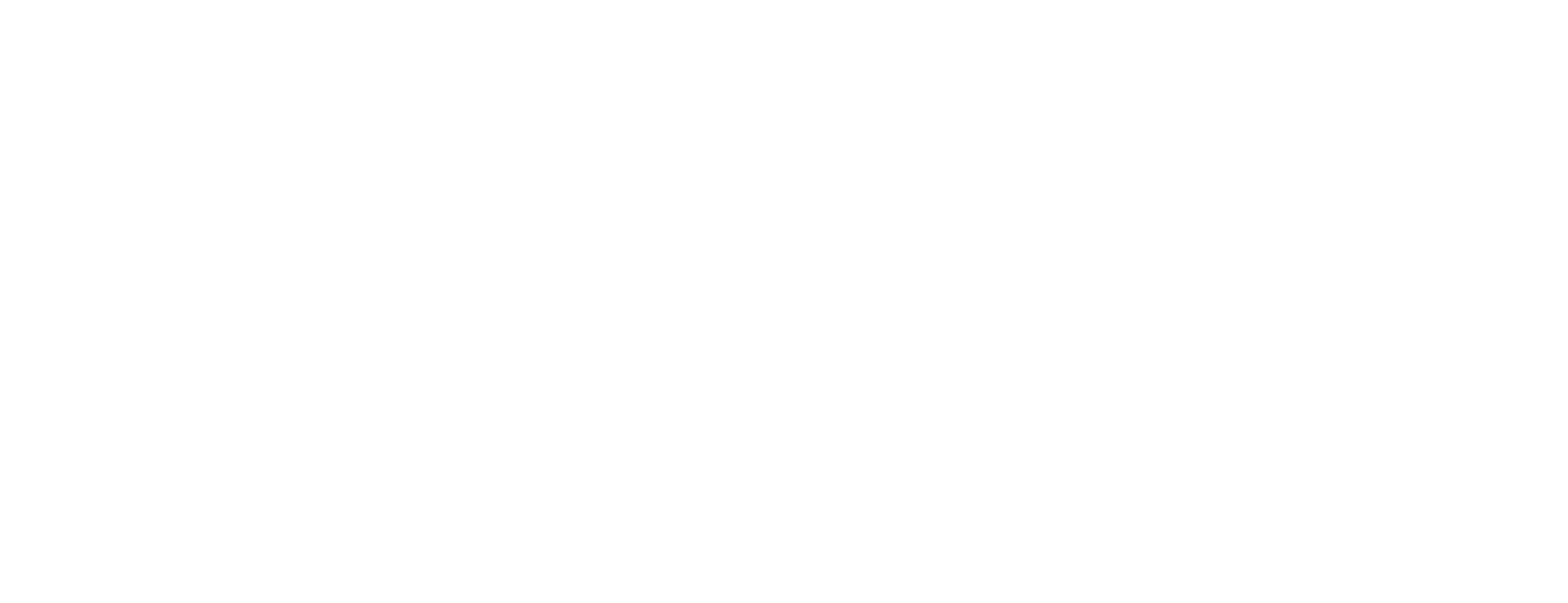}
\vspace{-4ex}
\caption{Influence of training depth (number of layers) on downstream classification task. The $x$ axis denotes the number of BGNN layers and the $y$ axis is the Micro $F_1$ score .}
\vspace{-2ex}
\label{fig:f1_score}
\end{figure}

\begin{table}[t]
    \centering
    \caption{Comparison of two layers end-to-end training with cascaded BGNN-Adv (Micro (Macro) $F_1$ score)}.
    \begin{tabular}{ccc}
        \toprule
         & {End-to-end training} & {Cascaded training}   \\
        \midrule
        Tencent & OOM & \textbf{0.622} \\
        Cora & 0.837 (0.809) & \textbf{0.859 (0.831)} \\
        Citeseer & 0.685 (0.642) & \textbf{0.768 (0.698)} \\
        PubMed & \textbf{0.859} (0.859) & 0.857 (\textbf{0.860)} \\
        \bottomrule
    \end{tabular}
    \label{table:cascaded_effective}
\end{table}

\textbf{Comparison with end-to-end training.} To compare cascaded training with end-to-end training, we specify an end-to-end training architecture in which there are multiple successive IDMP layers and one IDA output layer. In each IDMP layer, one domain's output, the hidden node vector, is the concatenation of raw features and another domain's aggregated node features obtained by IDMP operation. IDA is performed as the final output in an unsupervised manner. The results for comparison are shown in Table \ref{table:cascaded_effective}. We well-tuned the IDMP layer number for end-to-end training and find that the architecture with two IDMP layers achieves the best performance. The results demonstrate that our cascaded architecture achieves better performance on multiple datasets. In other words, this experiment verifies that cascaded training does not sacrifice statistical performance. We can also see from the results that end-to-end training runs out of memory (OOM) in the large-scale dataset Tencent. Since cascaded training also has system-wise benefits, such as faster training and lower memory cost (details are discussed in \textsection~\ref{subsec:system}), it is crucial for large-scale bipartite graphs.

\textbf{Effect of cascaded layer number.} With deeper cascaded training layers, the model can not only embed multi-hops information into the final representation but can also do so without expanding the memory cost and uncontrollable neighborhood expansion. The experiments in Fig. \ref{fig:f1_score} show the resultsa of Micro $F_1$ score of the downstream classification task along with the increase in the number of cascaded layers. Our observations and analysis are as follows:

\begin{itemize}
    \item Without cascaded architecture, a single layer BGNN has a relatively lower performance on all datasets. Particularly, a single layer BGNN means that there is no cascaded architecture during learning; only one-time optimization is performed. The reason is that a single layer only embeds one-hop neighbor information in the final representation. 
    \item Cascade-BGNN can achieve the best performance when there are two layers. Beyond two layers, the improvement is minimal, and even decreases slightly. This phenomenon is attributed to two reasons: 1. in the bipartite setting, nodes in two domains normally represent entirely different entities (e.g., user and community). Consequently, one node may not have enough of a correlation with its multi-hop connected node from the opposite domain. 2. Techniques that can train truly deep GNN are in demand. We suggest that future research may address this problem by incorporating/exploring the use of trained parameters from previous layers.
\end{itemize}

\subsection{Evaluation the efficiency of BGNN on large-scale Bipartite Graphs}
\label{subsec:system}

\begin{figure}[h]
  \centering
  \includegraphics[width=1.0\linewidth]{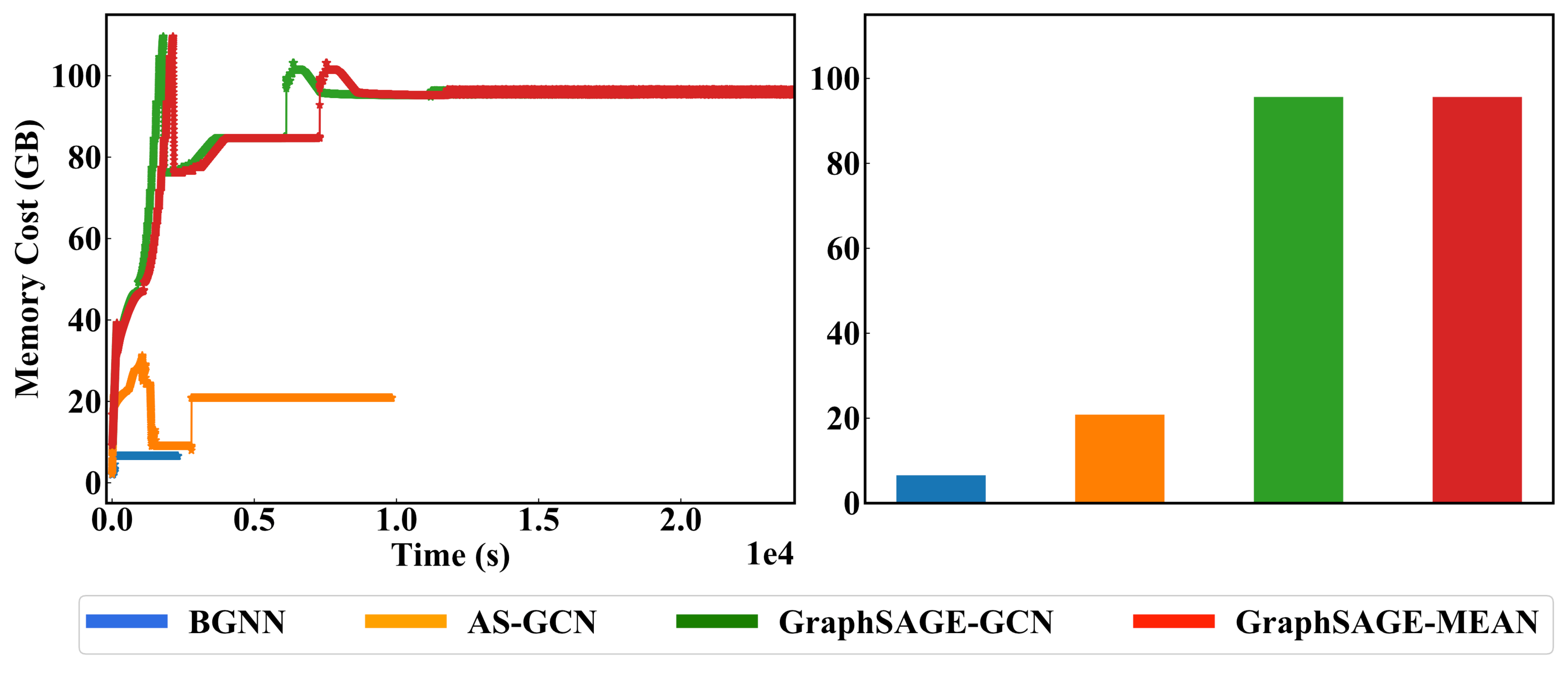}
\caption{Memory cost and training time on Tencent data. The $x$ axis in left figure denotes the wall-clock time in second, whereas the $y$ axis in both figures are the memory cost. The short blue line of BGNN and orange line of AS-GCN mean the training has finished, whereas the training time of GraphSAGE is too long to be shown.}
\label{fig:training_time}
\end{figure}
\textit{We developed a graph system based on Cascade-BGNN, which is introduced in the Appendix.}
In order to compare the scalability of BGNN against the baselines, we measure the training time and memory cost of algorithms on the Tencent large-scale dataset. As demonstrated in Fig. \ref{fig:training_time}, BGNN greatly outperforms baselines in terms of both space and time requirements. This final result is due to our experimental observations beforehand: (1) BGNN does not need to load the entire graph into memory--only one mini-batch is needed. However, all other methods require one to fill the graph into memory first, which corresponds to the huge increase in memory cost at the start of training. (2) BGNN does not require sampling in each graph convolutional layer and thus avoids an additional time-consuming procedure. (3) The unique unsupervised learning loss in BGNN based on adversarial learning does not require further computational processes. For example, in GraphSAGE, the unsupervised loss is based on random walks, causing it to increase significantly with the graph size. Additional nodes and a longer walk length are needed to maintain high performance, which consequently requires an even longer training time and a larger memory cost. (4) in \textsection~\ref{sec:cascaded}, we introduced that using cascaded training saves memory and avoids uncontrollable neighborhood expansion, reducing the memory cost and expediting training.

\section{Related work}
\label{sec:relatedWork}
Graph representation learning methods typically can be classified into two groups, supervised and unsupervised learning.

Unsupervised methods \cite{koren_matrix_2009} traditionally use the adjacency matrix to define and measure the similarity among nodes for graph embedding, which is referred to as matrix factorization. Other works explore using random walks on graphs to learn representation with the skip-gram model. DeepWalk \cite{perozzi_deepwalk:_2014} and Node2vec \cite{grover_node2vec:_2016} are typically representative of these methods to model homogeneous graphs. Some others extend this to heterogeneous graphs, where different nodes are in distinct feature domains, such as MethPath2Vec \cite{dong_metapath2vec:_2017} and PTE \cite{tang_pte_2015}. Although all these methods do not require node labels in their representation learning, they are shallow embedding approaches and have the following drawbacks. First, the nodes' features and graph structure are independent in the learning methods. Second, they are not efficient when applied to large- scale graphs. Third, the embeddings are transductive hence unseen nodes cannot be embedded with the model being learned so far.

As for supervised methods, the state-of-art graph-based neural networks have been used to learn node representation with the guide of node labels
\cite{kipf_semi-supervised_2016, schlichtkrull_modeling_2017}. In general, these methods perform a convolution by aggregating the neighbor nodes' information so that each node can learn a relationship between the nodes in the entire graph. However, these methods are task-specific: in another word, they require labels in downstream tasks to supervise the models. Some works \cite{hu2019pre, liu2019n, hamilton_inductive_2017, kipf_variational_2016} try to utilize GCN to do unsupervised learning on graphs by performing a random walk or matrix completion on the output of GCN embeddings. However, these approaches still face the same problems above. 
\section{Conclusion}
\label{sec:Conclusion}
In this paper, we propose Cascade-BGNN, a self-supervised node representation learning framework for bipartite graphs. Cascade-BGNN is domain-consistent, self-supervised, and efficient. Within Cascaded-BGNN, we propose Inter-domain Message Passing (IDMP) as the encoder and Intra-domain Alignment (IDA) by adversarial learning to address the node feature inconsistency issue. We further designed the cascaded training method to capture the multi-hop relationships in local bipartite structure, as well as to improve the scalability. Extensive experiments and theoretical analysis confirmed the effectiveness and efficiency of our models. 
 

\bibliography{KDD_2020}
\bibliographystyle{abbrv}
\clearpage
\appendix
\section{Appendix}

\begin{table*}[t]
\label{tab:cascaded}
  \caption{Reference for baselines code}
  \label{table:baseline}
  \centering
  \begin{tabular}{ll}
    \toprule
Baseline & Code link  \\
\midrule
Node2Vec (high performance version) & \url{https://github.com/snap-stanford/snap}  \\ 
VGAE & \url{https://github.com/tkipf/gae} \\
GraphSage & \url{https://github.com/williamleif/GraphSAGE}  \\
GCN  & \url{https://github.com/williamleif/GraphSAGE}  \\
AS-GCN & \url{https://github.com/huangwb/AS-GCN}\\
\bottomrule
  \end{tabular}
\end{table*}
\begin{table*}[t]
\label{tab:hyperparameters}
  \caption{Hyperparameters for BGNN on four datasets}
  \label{table:hyperparameters}
  \centering
  \begin{tabular}{c|ccccc}
    \toprule
& Hyperparameters & Tencent & Citeseer & Cora & PubMed \\
\midrule
BGNN-Adv & batch size & 600 & 400 & 400 & 700 \\ 
& epochs & 2 & 4 & 2 & 3 \\
& learning rate & 0.0004 & 0.0004 & 0.0004 & 0.0004 \\
& weight decay  & 0.0005 & 0.001 & 0.001 & 0.0005 \\
& dropout & 0.4 & 0.35 & 0.35 & 0.35 \\
& encoder output dimensions & 16 & 16 & 24 & 24 \\
    \midrule
BGNN-MLP & batch size & 500 & 64 & 128 & 128 \\ 
& epochs & 3 & 3 & 5 & 3 \\
& learning rate & 0.0003 & 0.001 & 0.001 & 0.0001 \\
& weight decay  & 0.001 & 0.0005 & 0.0008 & 0.005 \\
& dropout & 0.4 & 0.2 & 0.2 & 0.2 \\
& encoder output dimensions & 24 & 48 & 48 & 48 \\
& decoder hidden dimensions & 16 & 16 & 16 & 16 \\
  \bottomrule
  \end{tabular}
\end{table*}

This supplementary material provides source code for reproductivity, more details of the dataset, hyper-parameter settings, more experiment results, the infrastructure, and the future extension to large-scale bipartite graph system.

\section{Source Code for Reproductivity}
\textbf{Source code}. For experiment results reproductivity, we store this paper's source code at \textit{http://bit.ly/Cascade-BGNN}. Since we may refactor our code for further research, we maintain the original version of our code in this URL. We also provide the data that we use in this paper for running experiments. Besides the BGNN model, we also provide baseline codes we use in our experiments. Each model's code is organized in an independent directory. In order to help reproduce our results efficiently, in the \textit{README.md} file at the root directory, we organize a table of scripts for training procedure.

\section{Efficient Training System}

\begin{figure}[ht]
  \centering
  \includegraphics[width=0.51\textwidth]{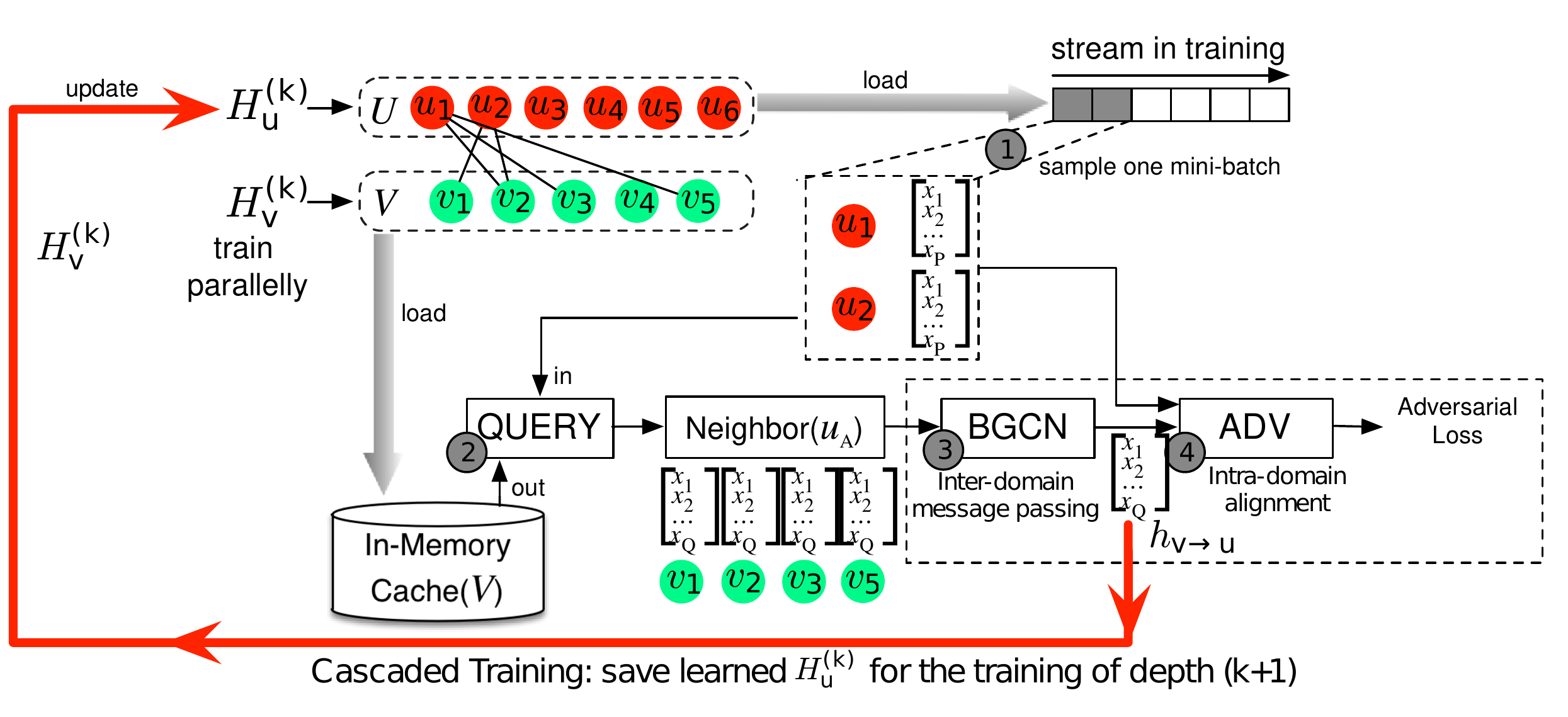}
  \caption{Cascaded Training Pipeline}
  \label{fig:trainingsystem}
\end{figure}

We depict a detailed diagram to illustrate our cascaded training pipeline from the perspective of set $U$, shown in Fig. \ref{fig:trainingsystem}. The step order is shown as a circle within a number. In step 1, we sample a mini-batch of node feature vectors from group $U$ (e.g., $u_{1}$ and $u_{2}$ with red color). In step 2, the QUERY operation takes the sampled node vectors as input and queries their neighbor node vectors from the opposite set $V$ (e.g, the queried neighbor vectors are $v_{1}$, $v_{2}$, $v_{3}$, and $v_{5}$ with green color). The inter-domain message passing is in step 3 where neighbor vectors are aggregated to $h_{v \rightarrow u}$. Then in step 4, the IDA (intra-domain alignment), taking sampled node vectors in $U$ and their aggregated neighbor vectors $h_{v \rightarrow u}$, as input is trained with an adversarial loss. After iterating all mini-batches with multiple epochs, the learned $H_{u}$ is saved for $(k+1)$th depth cascaded training. This pipeline is consistent with Algorithm 1.

\section{Proof of Theorem 1}
Here we give a proof of Theorem \ref{theorem_1}.
\begin{proof}
Note that for $D(P_1, P_2)$ measuring the distance of two distributions we have:
\begin{equation}
    D(P_1, P_3) \leq D(P_1, P_2) + D(P_2, P_3)
\end{equation}
We then have:
\begin{equation}
\begin{aligned}
    L_{M, u} - L_{M, v\rightarrow u} & = \sum_{h}p_u(h) \cdot D(P_u(y|h;\theta), \hat{P}(y|h))\\
    & - \sum_{h}p_{v\rightarrow u}(h) \cdot D(P_{v\rightarrow u}(y|h;\theta), \hat{P}(y|h))\\
    & \leq \sum_{h}(p_u - p_{v\rightarrow u})(h) \cdot D(P_{v\rightarrow u}(y|h;\theta), \hat{P}(y|h))\\
    & + \sum_{h}p_u(h) \cdot D(P_u(y|h;\theta), p_{v\rightarrow u}(y|h;\theta)) \\
    & = \epsilon L_{M, v\rightarrow u} + d
\end{aligned}
\end{equation}
\end{proof}

\section{More Experimental Evaluations}
In this section, we provide more information related to our paper, including detailed analysis of datasets and models implementation details.

\subsection{Model Implementation Details}

\textbf{Logistic regression}.  In order to evaluate our model output embedding performance, we use logistic regression to predict the nodes' label. We use the logistic \textit{SGDClassifier} from scikit-learn Python package. We split the nodes into 80 percentage for training and the rest for testing (30\% for validation).

\textbf{Model Implementation}. We use the code of baselines published by the author of the original paper. We summarize the baseline code we use in Table \ref{table:baseline}. We follow the parameter settings in their original papers and fine-tuned on our bipartite datasets. The Node2Vec is a high-performance version (C++), so its running time is comparable to ours. Since all the baselines are not designed for heterogeneous bipartite graph, in order to make a fair comparison with our models, we first transform the bipartite graph into a simple connected graph. We multiply the incidence matrix with its transpose to extract all two-hops connection. Since it is a bipartite graph, the two-hops connection of one set will only contain nodes in the exact same set. Through this simple transformation, the graph becomes to a single homogeneous graph, and all the baselines can achieve on it. 
\begin{figure}[h]
  \centering
{{\includegraphics[width=0.40\textwidth]{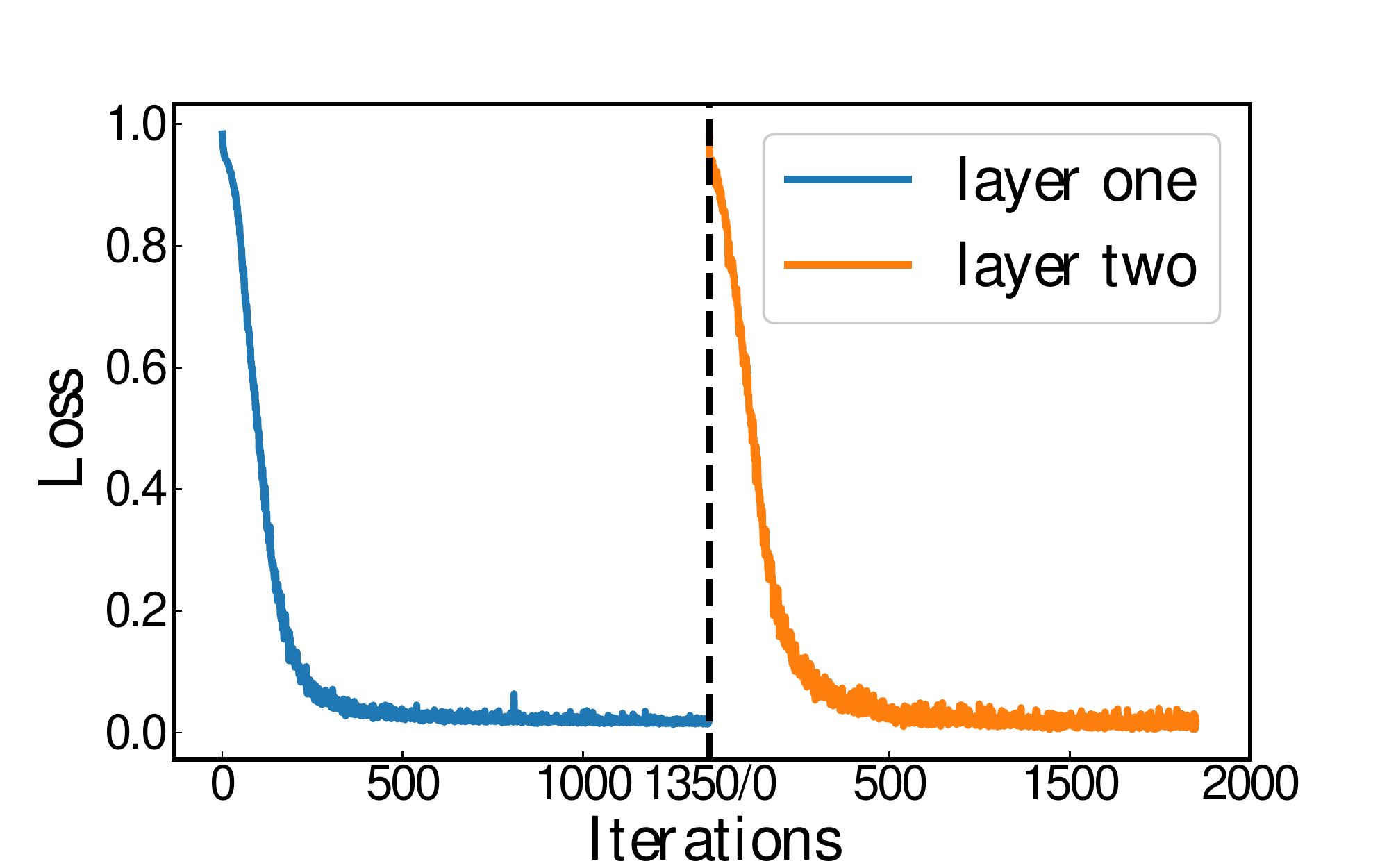}}}
\caption{BGNN-MLP training loss on Pubmed.}
\label{fig:mlp_loss}
\end{figure}

\textbf{Hyper-Parameters}.  We use a grid-search to tune our model on every dataset to find the best hyperparameters. Here, we list all the final hyperparameters of BGNN for different datasets.

As for epochs, we first search in a wide range and find that with small epochs size will achieve better performance. This also proves the reason why our model requires less training time. The BGNN-MLP model contains two dense layers with rectified activation layer and dropout layer in between. The output of the decoder is aligned in the range $[-1, 1]$ using hyperbolic tangent, which is the same distribution as the input features. As for BGNN-Adv model, the discriminator also contains two dense layers but with leaky rectified activation layer, which can avoid sparse gradient problem.

\subsection{More Experimental Results on Large-Scale Dataset}
\textbf{Training loss}.
Here we further show the training loss of BGNN-MLP model versus iterations on Pubmed dataset in Fig. \ref{fig:mlp_loss}. The reason is that PubMed is a medium-size dataset with balanced degree distribution, so it is a great dataset to illustrate. The loss function converges after around 500 iterations. The detail parameter settings can be found in table \ref{table:hyperparameters}.

\end{document}